\documentclass[aps,prl,showpacs,amssymb,reprint,nofootinbib]{revtex4-1}%
\usepackage{amsmath}%
\usepackage{amsfonts}%
\usepackage{amssymb}%
\usepackage{graphicx}
\usepackage{color}
\usepackage[english]{babel}
\usepackage{color}

\begin{document}

\title{Phase Separation on Bicontinuous Cubic Membranes: \\
Symmetry Breaking, Re-entrant and Domain Facetting} 
\author{Fabien Paillusson$^{*,\dagger}$, Matthew R. Pennington$^*$ and Halim Kusumaatmaja$^{*}$} 
\affiliation{$^*$Department of Physics, Durham University, South Road DH1 3LE \\ $^{\dagger}$School of Mathematics and Physics, University of Lincoln, Brayford Pool, LN6 7TS}
\pacs{} 
\date{\today}

\begin{abstract}
We study the phase separation of binary lipid mixtures that form bicontinuous cubic phases. The competition between non-uniform Gaussian membrane curvature and line tension leads to a very rich phase diagram, where we observe symmetry breaking of the membrane morphologies and re-entrant phenomena due to the formation of bridges between segregated domains. Upon increasing the line tension contribution, we also find facetting of lipid domains that we explain using a simple argument based on the symmetry of the underlying surface and topology. 
\end{abstract}

\maketitle

{\bf Introduction} - Lipid self-assembly can adopt an astonishing range of shapes and morphologies, from single bilayer structures to stacks and convoluted periodic structures \cite{Seddon}. Nature has, of course, exploited this polymorphism. A large number of organelles feature lipid-based structures, including synaptic vesicles, endoplasmic reticulum, and Golgi apparatus. At the same time, lipids are indispensable for detergency and foodstuffs industries \cite{Sagalowicz2010}, and membrane-based structures are increasingly exploited in biotechnological and biomedical applications, e.g. as efficient nanoporous scaffolds for tissue engineering \cite{Kapfer2011} or for gene silencing with siRNA \cite{Leal2010}.

In this letter we will focus on one particular type of mesophases that lipid mixtures in water can adopt, the so-called bicontinuous cubic phases (BCP) \cite{Templer1998,Angelov2003,Tyler2015}, whereby the lipids form a triply periodic lipid bilayer that separates two percolating and non-intersecting water channels  \cite{Benedicto97, Hyde12, Schoen12}. These phases have attracted attention due to their high surface area, continuity of the bilayer surface, and pore network. The amphiphilic nature of the lipids also allows other molecules to be embedded in them; for example, they have high propensity to enable membrane protein crystallization. Although the details remain unclear, it is thought that a combination of curvature induced phase separation on the cubic surface, a local destabilization of the cubic phase to a lamellar phase and a two dimensional reservoir of proteins provided by the cubic phase are responsible for the observed yield \cite{Caffrey15}. Here our interests are in the aforementioned curvature induced phase separation.

Both in the biological and synthetic systems, these lipid mesophases usually contain more than one lipid species. To the best of our knowledge, the distribution of different lipids across such a cubic surface, especially the possible demixing transitions under the  influence of {\it non-uniform} curvature of the membrane structures, is still not well-understood. Most studies on lipid phase separation focus on much simpler membrane geometries, such as lipid vesicles and supported membranes \cite{Veatch2003,Groves2006,Baumgart2007}. From a biological perspective, lateral lipid organizations into domains and membrane curvatures are ubiquitous features, and are known to play an important role for the membrane functionalities \cite{Brown1998,McMahon2005}. From a materials perspective, understanding the distribution of species of interest on a BCP may be the first steps towards a systematic and rational functionalization of BCPs, where active species can be localized into targeted domains. Finally, our work provides a comprehensive phase diagram, with predictions of distinguishing features which we hope will stimulate experimental verifications.

This letter is organized as follows. We first show that if the two species do not interact but induce different bending rigidities, then a single type of curvature induced phase separation occurs at all non zero area fractions. Upon considering interactions between the species, we observe a multiplicity of new modalities for the phase separation, including the formation of bridges between previously disconnected lipid domains. Moreover, we observe {\it facetting} of domains for which we provide a simple explanation relying on symmetry and topology.

\begin{figure}
  \includegraphics[width=0.98\linewidth]{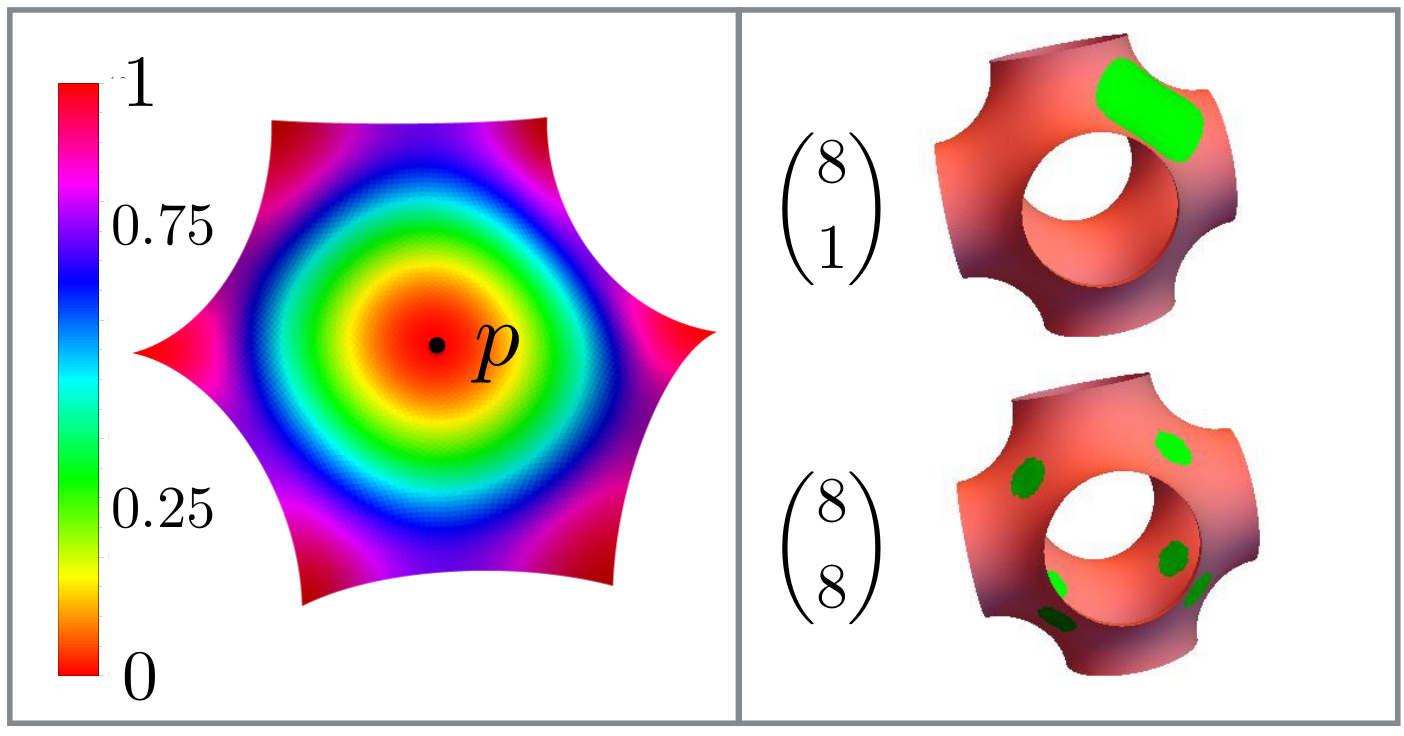} 
  \vspace{-2mm}
  \caption{\label{fig1} {\it Visualising the P-surface}. Left panel: normalised Gaussian curvature field $\frac{G(x)}{G_{min}}$ (note that $G(x) \leq 0$) on a single patch with the zero-curvature point $p$ at its centre. Right panel: curvature induced formation of A-lipid domains (in green) in $k$ patches among the 8 available denoted by $\binom 8 k$ (for $f_A = 0.07$). For the sake of illustration we show them for $k=1$ (top) and $k=8$ (bottom).}
\end{figure}

{\bf Segregation in absence of line tension} - In this paper we consider a binary lipid mixture or, alternatively, a mixture of lipids and proteins that has formed a minimal surface $S$ (with zero mean curvature everywhere), and ask what are the thermodynamically favoured repartitions of the species and how they depend on the bending rigidities and inter-species interactions. Here our focus is on the triply periodic surfaces which are known to be formed by lipid mixtures as well as mixtures of lipids and proteins in water \cite{Templer1998,Angelov2003,Tyler2015}.
We use standard notations P, D, and G for the {\it primitive}, {\it Diamond}, and {\it Gyroid} surfaces respectively. Since they are periodic, we characterize their properties per unit cell. 
To model a binary mixture on a curved surface, we use a straightforward extension of the Helfrich hamiltonian \cite{Lipowsky1991} to the case of a binary mixture on a minimal surface that reads \cite{SM, Brakke, Mathematica, Metropolis}:
\begin{equation}
H^{S}_{el}(f_A) = \delta \kappa \int_S d\mu_S(x) \: \sigma_A(x) G(x), \label{eq2} 
\end{equation}
where $x$ denotes a point on $S$, $d\mu_S(x)$ is the area measure on $S$ at $x$, $G(x) < 0$ is the gaussian curvature at point $x$, $f_A$ is the imposed area fraction of species $A$ (with $f_B = 1-f_A$), the field $\sigma_{A}(x) \in [0,1]$ is the {\it mean} occupation number of species $A$ at $x$, $\delta \kappa = \kappa_g^A-\kappa_g^B$ and $\kappa_g^{A,B}$ are the gaussian bending rigidities associated to the species $A$ and $B$ respectively. Our convention here is $\kappa_g^A < \kappa_g^B$, such that $\delta \kappa < 0$. It is also worth remarking that, since typically $\kappa_g < 0$, B domains are softer than A domains. Such a model may represent a coexistence between $L_o$ (liquid ordered; $A$-rich) and $L_d$ (liquid disordered; $B$-rich) domains, or alternatively, between lipid-rich and protein-rich domains.

%

To model the distribution $\sigma_A$ in a more tractable way, we first stress that any minimal surface with genus $g$ embedded in a flat torus $\mathbb{T}^3$ must contain $4(g-1)$ zero Gaussian curvature points \cite{Meeks90}. For the P, D and G surfaces, $g = 3$ and they have 8 zeros. Each zero is located at the centre of a hexagonal area we term as a {\it{patch}} (See Fig. \ref{fig1} here and Fig. 1 in \cite{SM}). The unit cell of either of the P, D or G surfaces can thus be partitioned into 8 equivalent patches $\{ \Sigma_i \}_{i=1..8}$ such that the unit cell surface $S = \cup_{i=1}^8 \Sigma_i$. 
We characterise the repartition of the lipids on $S$ by both the area fraction $f_A^i$ of lipid $A$ on each patch $\Sigma_i$ and the occupation number function $\sigma_A^i(x)$ in it. 
For each patch $\Sigma_i$, given $\sigma_A^i$, the entropy then reads $\mathbb{S}_i = -k_B  \int_{\Sigma_i}d\mu_S(x) [ \sigma_A^i \ln \sigma_A^i + (1-\sigma_A^i)\ln (1-\sigma_A^i)]$. Minimising the overall free energy $ \mathcal{F} = H^S_{el}(f_A)-T\sum_{i=1}^8 \mathbb{S}_i$ with respect to the occupation number for a given set of area fractions $\{ f_A^i \}_{i=1..8}$ leads to the typical Fermi-Dirac distribution $\sigma_A^{i*}(x) = [1+e^{-\beta \lambda_A^i + \beta \delta \kappa G(x)}]^{-1}$ where $\lambda_A^i$ is a Lagrange multiplier that imposes the value of $f_A^i$.



At low temperatures, the Fermi-Dirac distribution will reach a value close to unity for all points $x$ of $\Sigma_i$ with an energy lower than $\lambda_A^i$. 
The lowest energy point $p_i$ in a patch $\Sigma_i$ is the symmetry point of the patch which has exactly zero gaussian curvature (cf. Fig. \ref{fig1}).
Thus, at low $T$, the lipids $A$ will {\it fill} the neighbourhood of $p_i$ until they reach a critical {\it Fermi} curve $\mathcal{C}_F$, where $\{ x \in \mathcal{C}_F | \delta \kappa G(x) = \lambda_A^i \}$, beyond which there is no more lipids of type $A$ (cf. Fig. \ref{fig1}; a disconnected area occupied by lipid $A$ is termed as a domain). 
Close to $p_i$, one may use polar coordinates $(\rho_i, \theta_i)$ and, as a crude approximation, the space is assumed euclidean and circularly symmetric near $p_i$. This allows us to Taylor expand the function $G$ about $p_i$ up to the second order so that the curvature energy reads $H^{\Sigma_i}_{el}(f_A^i) \sim \delta \kappa \int_0^{R_i} 2\pi \rho_i d\rho_i \: [G''(p_i) \rho_i^2/2] \sim C (f_A^i)^2$, where $R_i$ is the mean radial distance of the Fermi curve from the point $p_i$ such that $f_A^i \approx \pi R_i^2 / \mu(\Sigma_i)$. Here $C$ is a constant and $\mu(\Sigma_i)$ is the area of the hexagonal patch $\Sigma_i$. Remarkably, in spite of the very crude approximations we have used, the predicted behaviour of the curvature energy $H^{\Sigma_i}_{el}(f_A^i) \propto (f_A^i)^{\alpha}$, with $\alpha = 2$, is close to what we observed in simulations for the P-surface where the exponent is found to be $\alpha = 1.83$ \cite{SM}. 

Next, upon minimising the total free energy $\mathcal{F}=C \sum_{i=1}^8 (f_A^i)^2$ with respect to the area fractions $f_A^i$ at fixed total area fraction $f_A = (\sum_{i=1}^8 f_A^i)/8$, it is easy to see that the ground state in repartition among the patches is always $f_A^{i} = f_A $ for all values of $f_A$, corresponding to the $\binom 8 8$ configuration in Fig. \ref{fig1}.


{\bf Effect of the line tension} - 
We have seen that, with only curvature, the $A$ lipids are evenly distributed among the 8 available patches and formed dense domains in the neighbourhood of zero curvature points at low temperature. This begs the question of how this picture changes if the $A-B$ interactions are not negligible {\it i.e.} if there are line tension effects arising with domain formation, which is a more realistic physical scenario. To answer this question, we now carry out computer simulations of the binary phase separation on the P-surface (the qualitative picture is the same for the D- and G-surfaces, as justified in \cite{SM}). There are several known approaches to model bicontinuous cubic membranes, from coarse-grained Molecular Dynamics simulations \cite{Marrink2009} to continuum field theoretical approaches \cite{Gompper1999,Lee2007,Yang2010}. Here we use Metropolis Monte Carlo simulations \cite{FrenkelBook} to resolve the thermodynamics of the system.

In our approach, we explicitly discretize a piece of the P-surface contained in a cubic cell. This can be efficiently done with the help of the Weierstrass-Enneper (W-E) representation of minimal surfaces \cite{MinimalSurfaces, Meeks90, Gandy-P}. 
Upon discretization, the binary mixture can then be modelled as an Ising-like problem (see \cite{SM} for technical details). A spin variable $s$ is associated to each site and takes either value $0$ (for species $B$) or $1$ (for species $A$). The curvature hamiltonian of Eq. \eqref{eq2} thus maps exactly onto a system of magnetic spins on a network $\mathcal{N}(S)$ with a node-dependent external magnetic field and reads:
\begin{equation}
H^S_{el}(f_A) = \sum_{i \in \mathcal{N}(S)} \delta A_i \delta \kappa G_i s_i, \label{eq4}
\end{equation}
where $\delta A_i$ is the area of the tile $i$ on the surface. In this language, at any finite $f_A$, species $A$ (spin variable $s = 1$) will occupy sites with the lowest value of $\delta A_i \delta \kappa G_i$ to minimize the total energy, as we have analyzed with a different vocabulary in the previous section. To model the $A-B$ interspecies interactions, we choose a short-range nearest neighbours interaction which directly translates into the line tension of the lipid domains:
\begin{equation}
H^S_{A-B}(f_A) \equiv J \sum_{i \in \mathcal{N}(S)} \sum_{j \in \langle i \rangle} (s_i+s_j-2s_is_j) \delta L_{ij} \: , \label{eq5}
\end{equation}where $J$ sets the magnitude of the exchange interactions, $\delta L_{ij}$ is the length of the edge shared by cells $i$ and $j$, and $(s_i+s_j-2s_is_j)=1$ when $s_i \neq s_j$ and 0 otherwise. 

\begin{figure}
 \vspace{2mm}
  \includegraphics[width=0.99\columnwidth]{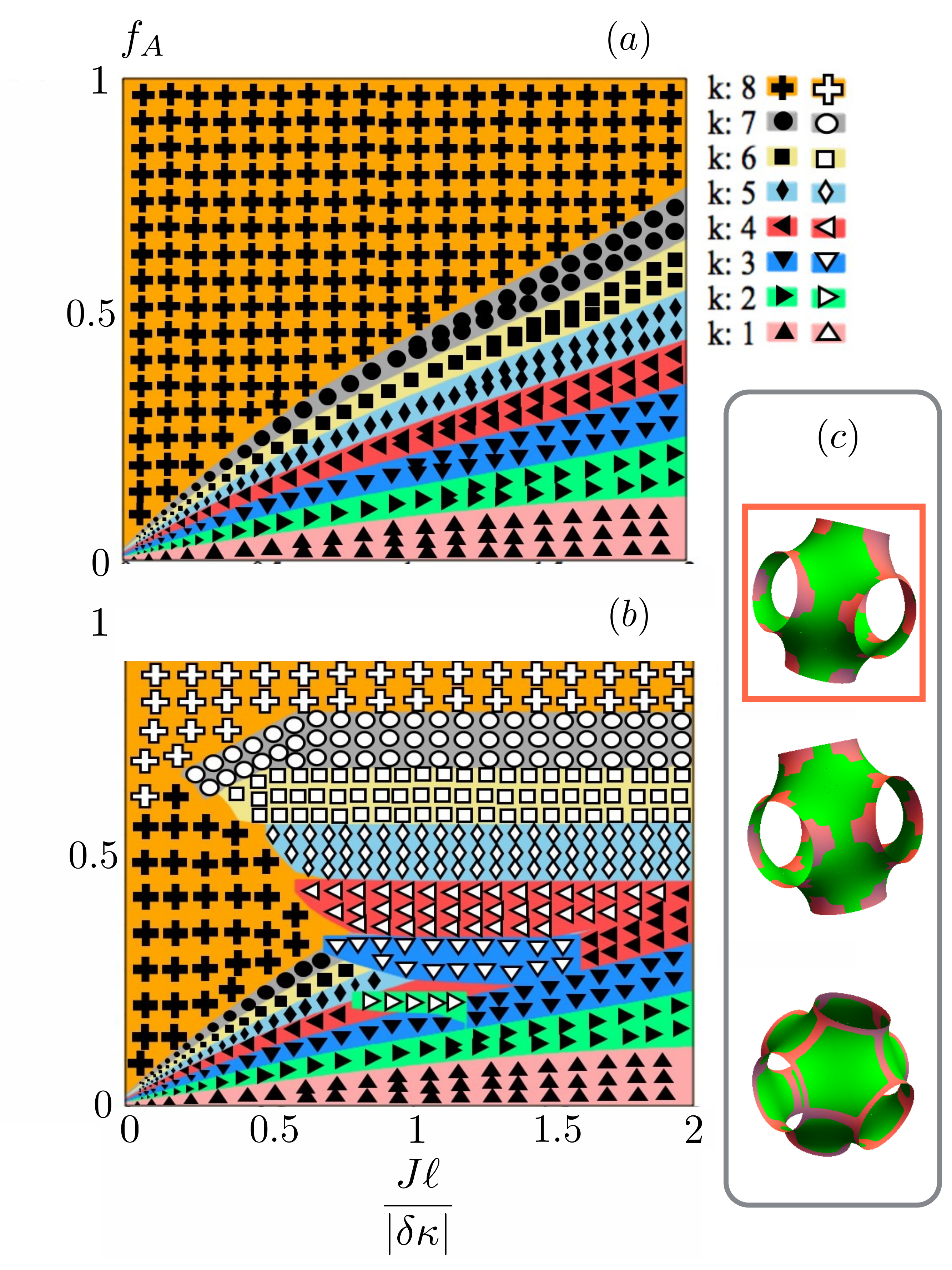}
  \vspace{-4mm}  
    \caption{\label{fig2}
    {\it Phase diagrams} of the number of patches $k$ occupied by lipid $A$ domains among the $8$ available as a function of the area fraction $f_A$ and the ratio $J \ell /|\delta \kappa|$. $\ell$ is the lattice spacing of the cubic unit cell.
For clarity, we show the phase diagrams $(a)$ excluding (filled symbols) and $(b)$ including (open symbols) the possibilities of bridge formations between lipid $A$ domains. In $(c)$ are shown configurations with increasing energy from top to bottom for $f_A = 0.75$ and $J \ell/|\delta \kappa| = 0.5$. The boxed configuration with 7 connected domains corresponds to the most stable phase.
    }
\end{figure}

{\it Symmetry breaking} - As it is evident, the hamiltonian in Eq. \eqref{eq5} is equivalent to an Ising model of ferromagnetism and therefore should lead to the same phenomenology: above a critical temperature $T^*(J)$, the system is paramagnetic and the two lipid species are mixed; while below $T^*$, the system becomes ferromagnetic and a {\it symmetry breaking} favouring "lumping" of spins in spatial regions (segregation) occurs. There is, however, one crucial difference between the standard Ising model and our model. For the former, line tension effects always dominate demixing: domains of $A$ lipids coalesce to minimize the overall interfacial energy. In our model, this coalescence mechanism competes with the curvature-induced mechanism described in the previous section. 

The first effect of line tension is to re-shuffle the (energy) ranking of configurations $\binom 8k$ with $k$ patches occupied by the $A$ species by shifting $down$ the low $k$ configurations (because they have a lower interfacial cost) and $up$ the high $k$ ones (because they have a high interfacial cost). A first account of the competition between curvature and line tension consists in assuming that the total energy of a configuration $\binom 8 k$ at a given packing fraction $f_A$ would read $k[H^{\Sigma}_{el}(8f_A/k)+H^{\Sigma}_{A-B}(8f_A/k)]$ {\it i.e.} as the sum of the free energy of individual patches of equal size. This {\it summation} approximation is valid when isolated domains are formed at the centre of the hexagonal patches, and one finds that increasing $f_A$ at fixed $J \ell /|\delta \kappa|$ always favours, eventually, higher $k$ values, in agreement with Monte Carlo simulation results shown in the phase diagram in Fig. \ref{fig2}$(a)$.

The caveat is that this summation approach is only valid when the $A$ species domains are disconnected. Above certain $f_A$ values, the lowest energy configurations are in fact those in which domains of lipid $A$ span across multiple patches (see Fig. \ref{fig2}). These {\it bridges} between patches essentially make the domains interact negatively and in a non-pairwise fashion. 
The location of these bridges coincides with the lowest curvature energy regions at the patch boundary (c.f. Fig. \ref{fig1}).
Taking these configurations into account, the phase diagram in Fig. \ref{fig2}$(b)$ shows that the simple picture of Fig. \ref{fig2}$(a)$ only holds for small $J \ell/|\delta \kappa|$ and $f_A$. In fact, one observes {\it re-entrant} behaviours whereby a configuration $\binom 8 k$ previously unfavored in the disconnected regime, becomes re-favored thermodynamically. We note that when bridges are formed in the $\binom 8 8$ configuration (open crosses in Fig. \ref{fig2}$(b)$), the segregated and continuous phases are effectively {\it inverted} (lipid $B$ domains are surrounded by $A$).

\begin{figure}
 \vspace{2mm}
   \includegraphics[width=0.48\linewidth]{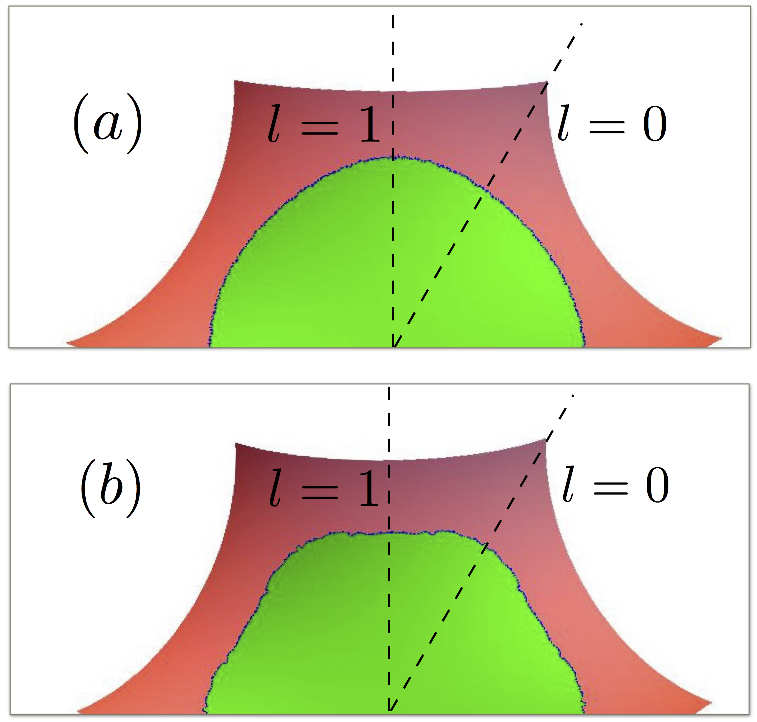} 
   \includegraphics[width=0.5\linewidth]{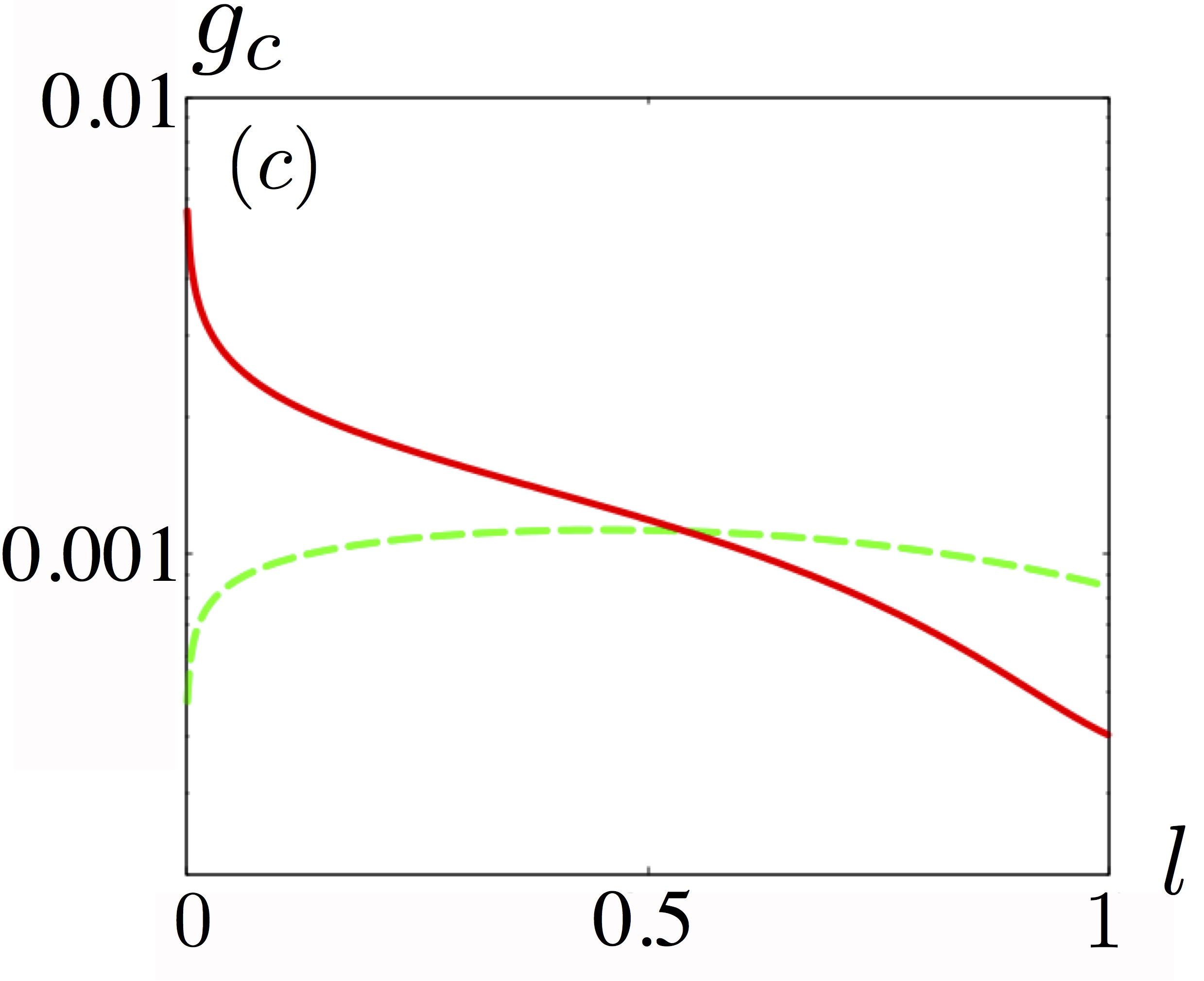} 
    \vspace{-2mm}
  \caption{\label{fig3} {\it Domain facetting.} Shape of a lipid $A$ domain in the neighbourhood of the zero-curvature point $p$ of a single hexagonal patch. Note that only a half of the patch is represented. $(a)$ in absence of line tension and $(b)$ with high line tension ($J \ell/|\delta \kappa| > 1$). $(c)$ Geodesic curvature as a function of the curvilinear coordinate $l$ in absence of line tension (dashed) and with high line tension (solid).}
\end{figure}

{\it Domain facetting} - 
Another distinguishing feature that appears with line tension is the {\it facetting} of the domains formed by the $A$ lipids. This effect is shown in Fig. \ref{fig3}$(b)$ where the domain almost draws a hexagon compared to Fig. \ref{fig3}$(a)$ where the shape is more rounded, thus the term ``facetting''. To explain this, we recall that in general if the underlying manifold has an n-fold rotational symmetry, we expect the bounding curve that minimises the perimeter length of a domain with fixed area to be a regular n-gon whose sides are geodesics of the underlying manifold. 
Moreover, on an anisotropic curved surface, not all orientations of a regular $n$-gon are equivalent as they lead in principle to different total perimeter lengths. Thus, we interpret the bounding curve in Fig. \ref{fig3} (b) with 6-fold symmetry
to be the curve that minimises both shape and orientation at the same time.

To test the above rationale, we estimate the geodesic curvature along the bounding curves of the two representative examples shown in Fig. \ref{fig3}$(a)$ and $(b)$ (cf. {\it e.g.} Ref. \cite{Kamien02}). In these figures, the curvilinear coordinate $l \in [0,1]$ is the normalised arc length of each curve which enables the comparison of the geodesic curvature for curves with different total lengths in Fig. \ref{fig3}$(c)$. In absence of line tension, the geodesic curvature $g_c$ is approximately constant around the boundary. With a large line tension, $g_c$ reaches very high values for $l$ close to zero, but is much smaller than that without line tension as $l$ approaches $1$. This is consistent with the above explanation although it shows that the facetting is not perfect. It nevertheless sheds light on what happens as we approach the ideal facetting case: the geodesic vanishes almost everywhere except close to $l \simeq 0$ where it diverges. This divergence is representative of the wedge formed by the intersection of two geodesics of the 6-gon and whose angle $\gamma$ can be estimated to be $\gamma = 2\pi/3 - (6 |\delta \kappa|)^{-1}H^{\Sigma_i}_{el}(f_A^i)$ for an area $f_A^i \mu(\Sigma_i)$ \cite{SM}.

{\bf Discussions} - Let us start by estimating where typical lipid mixtures are located in the phase diagram of Fig. \ref{fig2}. For a mixture of DOPC, sphingomyelin, and cholesterol forming coexisting $L_o$ and $L_d$ domains, $J \simeq 1.2$ pN and $|\delta \kappa| \simeq 3 \times 10^{-19}$ J \cite{Semrau,Baumgart}. Most synthetic BCPs, however, are formed using the lipid Monoolein, which is known to have a low bending rigidity with $|\kappa_G| \sim \kappa_m < 10 k_BT$ \cite{Shearman}. Here $\kappa_m$ is the (mean curvature) bending rigidity. Using these values and taking the typical lattice spacing of a BCP $\ell \simeq 10-100$ nm \cite{Barriga,Almsherqi,Deng}, this leads to $J\ell/|\delta \kappa|$ in the range of $O(0.1)$-$O(1)$ considered in this paper.

When the line tension effects can be neglected, the natural curvature of the surface alone is enough to a) induce segregation in all surfaces and b) the segregation is such that domains form in the same proportions on all available patches on the surface. We then confirmed this theoretical prediction by numerical calculations on the P-surface and looked at the effects of non-zero line tension. The latter gives rise to two important features. (i) Below the demixing critical temperature, it favours the formation of bigger domains in a fewer number of patches available on the surface that we characterize with a corresponding phase diagram. We also observe re-entrances in this patch-occupation space due to the formation of bridges between domains on neighbouring patches. Some of these morphologies should lead to distinguishing features ({\it e.g.} different x-ray scattering signatures due to the change in symmetry), and we hope this work will stimulate experimental works to verify our predictions. (ii) In the large line tension limit, we observed a facetting of the domains for which we provided a simple explanation and that we can relate to the curvature energy of a domain on a patch. 

Predicting patterning on cubic membranes is the vital first step towards their systematic and rational functionalization. On one hand, the ability to localize molecular species by design into targeted domains can be beneficial for controlled release in drug delivery or of chemical substances \cite{Rizwan,Garti}, and for templating self-assembly \cite{Brujic} or phase separation \cite{Boltau1998, Xue2012} in the surrounding fluids. On the other hand, suppressing phase separation between lipid species or between lipids and proteins can be desirable in applications such as protein crystallization \cite{Caffrey15}, where segregation at an incorrect stage can strongly hamper the efficiency of the applications.

There are also a number of avenues for future work. Firstly, here we have assumed that the BCP remains a minimal surface. 
A closer inspection based on the theory developed in \cite{Lipowsky1992} for domain-induced  budding shows that the conclusions presented here can be qualitatively affected when $ \kappa_m/|\delta \kappa| < 0.4$ \cite{SM}. However, estimates of this ratio for a wide range of lipid bilayers and monolayers in the literature show that it is only rarely below 1 \cite{Hu12}. This suggests that the minimal surface assumption is very reasonable for realistic parameter values. Further work is however still needed to fully assess how membrane deformation, including budding instability, affects the phase diagram of multicomponent BCPs.
Secondly, the present work tacitly assumes that the membrane domains are formed by lipids of the same species in the two leaflets ({\it registration} phase). Indeed, recent work on flat bilayers suggests that registered domains is the thermodynamically favoured phase for a wide range of lipid mixtures \cite{Olmsted2015}. It would be interesting to relax this assumption to probe how curvature affects registration/anti-registration and how, in turn, registration/anti-registeration may affect the bilayer morphology. Thirdly, the system considered here provides an excellent setup to study how non-uniform curvature may affect the nature of the demixing phase transition.

{\bf Acknowledgements} - We thank J. M. Seddon, P. D. Olmsted, J. J. Williamson, N. J. Brooks, D. Frenkel, F. Schmid and C. Semprebon for useful discussions,
and the Biophysical Sciences Institute in Durham for a summer studentship for MRP. This work is funded by EPSRC (EP/J017566/1).


\newpage 

\clearpage

\begin{center}
\textbf{\large Supplemental Materials}
\end{center}

\section{Free energy of a binary mixture on a membrane}
\subsection{Free energy for single species membranes}
For a membrane comprising of a single specie, the free energy of a given shape $S$ can be described via the Helfrich hamiltonian \cite{Lipowsky1991}:
\begin{align}
H_{el} = \int_{S} d\mu_S \: [\kappa_m (C- C_{sp})^2 + \kappa_g G], \tag{SM1} \label{Hel1} 
\end{align}where $C \equiv (c_1 + c_2)/2$ is the local mean curvature, $G = c_1 c_2$ is the local gaussian curvature, $c_{1,2}$ are the local principal curvatures, $C_{sp}$ is the intrinsic mean curvature of  the surface and $\kappa_{m,g}$ are the mean and gaussian bending rigidities respectively. Furthermore, the Gauss-Bonnet theorem states that \cite{Kamien02}:
\begin{equation}
\int_S d\mu_S \:G + \oint_{\partial S} dl \:g_c = 2\pi \chi(S),  \tag{SM2 }\label{Hel2}
\end{equation}where $\partial S$ stands for the boundary of $S$, $g_c$ for the local geodesic curvature of the boundary \cite{Kamien02} and $\chi(S)$ for the Euler characteristic of $S$. The key consequence is that, if the boundary and the topologies of the problem are fixed, minimizing Eq. \eqref{Hel1} is equivalent to minimizing only the integral over the mean curvature $C$. In the absence of intrinsic curvature, the lowest energy solutions correspond to surfaces with exactly $C = 0$ at all points: such surfaces are called {\it minimal} surfaces \cite{MinimalSurfaces}. 
Here we focus on the triply periodic surfaces which are known to be formed by lipid mixtures in water \cite{Templer1998,Angelov2003,Tyler2015}.
We use standard notations P, D, and G for the {\it primitive}, {\it Diamond}, and {\it Gyroid} surfaces respectively. Since they are periodic, we characterize their topology with their Euler characteristic per unit cell ({\it e.g.} in Eq. \eqref{Hel2}). 

\subsection{Free energy for a binary mixture on a minimal surface}
If instead of a single species, the membrane comprises two different species, then different allowed mixture configurations may have different topologies and one cannot disregard anymore the gaussian curvature contribution to the energy. The simplest extension of Eq. \eqref{Hel1} to a binary mixture would then read:
\begin{align}
\tilde{H}^S_{el}(f_A) = \int_{S} d\mu_S(x) \: [\kappa_g^A \sigma_A(x) + \kappa_g^B(1-\sigma_A(x))]G(x) \tag{SM3} \label{Hel3}
\end{align}where $x$ denotes a point on $S$, $d\mu_S(x)$ is the area measure on $S$ at $x$, $G(x)$ is the gaussian curvature at point $x$, $f_A$ is the imposed area fraction of species $A$ (with $f_B = 1-f_A$), the field $\sigma_{A}(x) \in [0,1]$ is the {\it mean} occupation number of species $A$ at $x$ and $\kappa_g^{A,B}$ are the gaussian bending rigidities associated to the species $A$ and $B$ respectively. We then choose the free energy $\tilde{H}^S_{el}(f_A = 0)$ as reference so that, in practice, we look at the free energy $H^S_{el}(f_A ) \equiv \tilde{H}^S_{el}(f_A) - \tilde{H}^S_{el}(f_A = 0)$ which yields:
\begin{align}
H^S_{el}(f_A) = \delta \kappa \int_{S} d\mu_S(x) \:  \sigma_A(x) G(x) \tag{SM4} \label{Hel4}
\end{align}where $\delta \kappa = \kappa_g^A - \kappa_g^B$. Eq. \eqref{Hel4} is the starting point of our study.

\section{Weierstrass-Enneper representation}
\subsection{General formulation}
It can be shown that, locally, any minimal surface can be conformally mapped onto the complex plane via the Weierstrass-Enneper (W-E) representation \cite{MinimalSurfaces}. More precisely, the W-E is a map from $\mathbb{C}$ to $\mathbb{R}^3$ which, to a point $(u,v)$ in an open subset of $\mathbb{C}$, uniquely associates a point $(x(u,v), y(u,v), z(u,v))$ of $\mathbb{R}^3$ that belongs to a minimal surface via:

\begin{align}
&& x(u,v) = \Re \left\lbrace \int_{w_0}^{u+iv} dw \:f(w)(1-g(w)^2) \right\rbrace  \tag{SM5}\label{WE1} \\
&& y(u,v) = \Re \left\lbrace \int_{w_0}^{u+iv} dw \:if(w)(1+g(w)^2) \right\rbrace  \tag{SM6}\label{WE2} \\
&& z(u,v) = \Re \left\lbrace \int_{w_0}^{u+iv} dw \:2f(w)g(w) \right\rbrace  \tag{SM7}\label{WE3} 
\end{align}where $g$ is a holomorphic function and $f$ is a meromorphic function such that $fg^2$ is analytic.

\vspace{2mm}

The gaussian curvature $G(x,y,z)$ at any point of a surface represented by Eqs. \eqref{WE1}, \eqref{WE2} and \eqref{WE3} can be expressed as a function of $w=u+iv$ via:

\begin{equation}
G(w) = - \left(\frac{4|g'(w)|}{|f(w)|(1+|g(w)|^2)^2} \right)^2 .  \tag{SM8}\label{WE4} 
\end{equation}
The negative sign is characteristic of minimal surfaces. Since the mean curvature $(c_1 + c_2)/2 $ is zero everywhere, it implies that the gaussian curvature is always negative or zero. The W-E being a conformal map, it preserves the angles. Distances, however, are not conserved when mapping an infinitesimal segment from $\mathbb{C}$ to $\mathbb{R}^3$ and are scaled by a factor $\Lambda(w)$ given by
\begin{equation}
\Lambda(w) = \frac{|f(w)|(1+|g(w)|^2)}{2} \tag{SM9} \label{WE5}
\end{equation} 

\subsection{Triply periodic Schwartz surfaces}
The three Schwartz surfaces {\it P}, {\it D} and {\it G} can be obtained by choosing:
\begin{align}
&& g(w) = w \tag{SM10}\label{WE6} \\
&& f(w) = \frac{e^{i \theta_B}}{\sqrt{w^8-14w^4+1}} \tag{SM11}\label{WE6bis}
\end{align}where $\theta_B$ is the Bonnet angle such that $\theta_B = 0 $ for the {\it D} surface, $\theta_B = \pi/2$ for the {\it P} surface and $\theta_B = \mathrm{cotan}(K(1/4)/K(3/4))$ for the {\it G} surface. $K$ is a complete elliptic integral of the first kind. Since $e^{i \theta_B}$ only changes the phase in the W-E representation, this means that the {\it P}, {\it D} and {\it G} surfaces are simply related by an isometry called the Bonnet transformation and share many of their physical properties.

\subsection{Independence of the free energy on the member of the Bonnet family}
In particular, having the W-E map in mind, the whole integral in Eq. \eqref{Hel4} can be thought of as an integral in the complex plane. Using the fact that $d\mu_S(x(w)) =\Lambda^2(w)dudv$, the Eq. \eqref{Hel4} can be recast as:
\begin{equation}
H^S_{el}(f_A) = \delta \kappa \int_{\mathcal{A}(S)} dudv \: \Lambda^2(w) \: \sigma_A(w) G(w) \tag{SM12}\label{WE8} 
\end{equation}where $\mathcal{A}(S)$ denotes the atlas used in $\mathbb{C}$ to characterize $S$. It is worth noting that in the integrand of Eq. \eqref{WE8}, the W-E functions $f$ and $g$ only appear via their complex modulus and therefore their contribution to the curvature energy would be unchanged by a phase factor. Thus, we concludethat the curvature energy of a binary mixture on a minimal surface is independent of which member of the Bonnet family is considered and, in particular, so is its ground state. In a similar fashion, a continuous model of the line tension contribution would read formally:
\begin{equation}
H_{A-B}(f_A) = J \int_{\mathcal{I}(A-B)} d\mu_l(x) \tag{SM13}\label{WE9}
\end{equation}where $\mathcal{I}(A-B)$ denotes the set of points belonging to the $A-B$ interface on $S$ and $d\mu_l(x)$ the length measure on $S$. Again, the integral can be thought as an integral on $\mathbb{C}$ by virtue of the W-E map. Furthermore, the length measure on $S$ can be expressed in term of the length measure in $\mathbb{C}$ via $d\mu_l(x(w)) = \Lambda(w) |dw|$. Eq. \eqref{WE9} can thus be rewritten as:
\begin{equation}
H_{A-B}(f_A) = J \int_{WE^{-1}(\mathcal{I}(A-B))} |dw| \: \Lambda(w) . \tag{SM14}\label{WE10}
\end{equation}
\begin{figure}
 \vspace{2mm}
  \includegraphics[width=0.49\columnwidth]{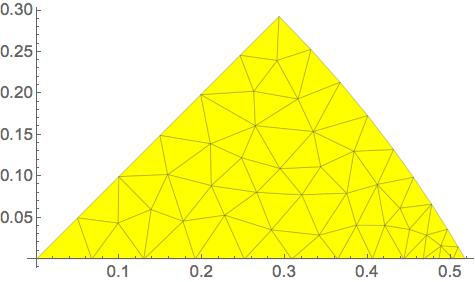} 
   \includegraphics[width=0.49\columnwidth]{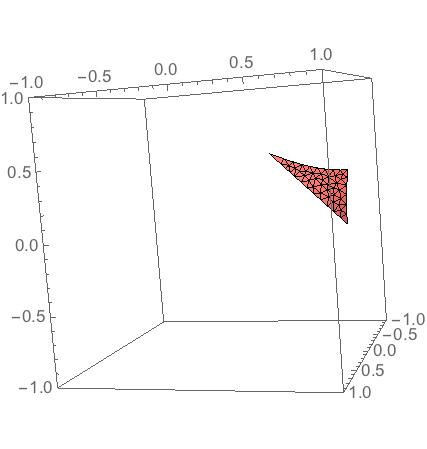}\\
    \includegraphics[width=0.49\columnwidth]{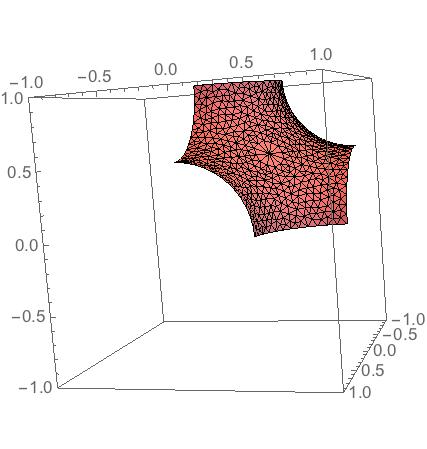} 
     \includegraphics[width=0.49\columnwidth]{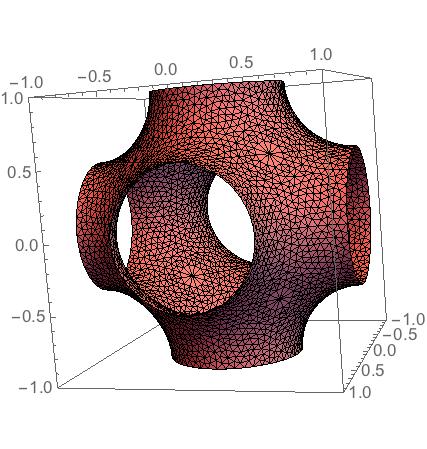}  
  
  \caption{\label{fig1} {\it Constructing the P-surface}. Top left: complex domain of the fundamental patch. Top right: fundamental patch in three dimensions. Bottom left: hexagonal patch ($\Sigma$ in the main article) made of 12 fundamental patches. Bottom right: Full P-surface per cubic unit cell $S$ made of 8 patches $\Sigma$.  }
\end{figure}

As before, the W-E functions $f$ and $g$ only contribute to the integral via their complex modulus and therefore $H_{A-B}$ is independent of the member of the Bonnet family under study.
As a consequence, the phenomenology of symmetry breaking and bridging-induced reentrant behaviour depicted in the phase diagrams of the P-surface in Fig. 2 of the main text hold in fact for all three Bonnet surfaces. The details of the phase diagram may slightly differ, however, as the cubic cells of the {\it G} and {\it D} surfaces do not contain the same surface area as the {\it P} surface. This is outside the scope of this letter, and we will discuss these details in a separate publication.

\subsection{Angle of intersection between two geodesics}
The Euler characteristic of a compact domain $\mathcal{D}$ of $A$ lipids that is not bridged to another domain on a neighbouring patch is $1$. Applying the Gauss-Bonnet theorem (Eq. \eqref{Hel2}) to such a domain gives:
\begin{equation}
\int_{\mathcal{D}} d\mu_S(x) \: G(x) + \oint_{\partial D} dl \: g_c(x) = 2\pi . \tag{SM15} \label{GB1} 
\end{equation}Up to a factor the first term of the {\it l.h.s} of Eq. \eqref{GB1} is simply the curvature energy of the domain that we can denote $H^{\Sigma_i}_{el}(f_i)$ in referring to notations introduced in the main text.
By using the fact that a patch has a 6-fold rotation symmetry, we can split the boundary integral of the geodesic curvature into 6 equivalent parts. If each piece of the boundary is almost everywhere a geodesic, then the second term of the {\it l.h.s} of Eq. \eqref{GB1} has zero integrand everywhere except at points where the geodesics meet. The total value of the contour integral becomes simply a sum over intersection angles that we call $\theta$. We thus have:
\begin{equation}
\theta = \frac{\pi}{3} + \frac{H^{\Sigma_i}_{el}(f_A^i)}{6 |\delta \kappa|}. \tag{SM16} \label{GB2}
\end{equation}Finally, the actual interior angle $\gamma$ between two geodesics making the hexagonal-like facetted domain is in fact the complementary angle of $\theta$ and reads:
\begin{equation}
\gamma = \frac{2\pi}{3} - \frac{H^{\Sigma_i}_{el}(f_A^i)}{6 |\delta \kappa|}. \tag{SM17} \label{GB3}
\end{equation}
Note that in the case where the curvature energy vanishes but the symmetry is still imposed, we retrieve the interior angle of a planar hexagon as expected.

\section{Numerical modelling of the P-surface}
\subsection{Fundamental patch}
All well behaved minimal surfaces admit a description in terms of a fundamental patch in $\mathbb{R}^3$ that is repeated by using the symmetries of the surface. By the W-E representation, this fundamental patch is associated to a fundamental domain of the complex plane. For the {\it P}, {\it D} and {\it G} surfaces, the fundamental domain is the set of complex points with positive real part bounded by the lines along the vectors $(1+i)/\sqrt{2}$ and $1$ and by the circle of radius $\sqrt{2}$ whose center is located at the point $-(1+i)/\sqrt{2}$. 

The top left of Fig. \ref{fig1} shows the fundamental domain in $\mathbb{C}$. The triangular tessellation is obtained by using the {\it Surface Evolver} package \cite{Brakke} and the images plus the management of the network structure have been performed with the {\it Mathematica} software \cite{Mathematica}. For the sake of illustration, Fig. \ref{fig1} shows a coarse tessellation of the fundamental domain. The tessellations we used in the paper are typically 100 times finer. By using Eqs. \eqref{WE1}-\eqref{WE3} and \eqref{WE6} and \eqref{WE6bis}, we get a three dimensional realization of the fundamental patch that is represented in the top right of Fig. \ref{fig1}. Then, following Ref. \cite{Gandy-P}, we can generate first a full hexagonal patch of the P-surface ($\Sigma$) by replicating and stitching together 12 fundamental patches as seen in the bottom left of Fig. \ref{fig1}. The full cubic cell representation of the surface $S$ is then obtained by combining 8 such hexagonal patches with the right symmetry operations as illustrated on the bottom right of Fig. \ref{fig1}. A similar procedure, albeit with different arrangements of the fundamental patches, can also be carried out for the D- and G-surfaces. 

\subsection{Monte Carlo simulations}
As emphasized in Eq. \eqref{WE8}, the curvature energy can be recast in terms of a sum over points on a euclidean (complex) plane of a curvature field that multiplies a scaling field. Moreover, the discretized surface $S$ on the bottom right of Fig. \ref{fig1} is made of a network of cells $\mathcal{N}(S)$ which are either an original version or a replica of a cell in the fundamental patch. Thus, the whole set of values of the curvature and scaling fields on the whole network is determined solely by that of the sub-network of cells in the fundamental patch. The particular topology of the P-surface (of genus 3 in a cubic cell) is then accounted for by the topology of the network {\it i.e.} by assigning the right neighbours to each cell. If we add a species field $s$ into the picture such that $s_i= 1$ if cell $i \:\in \: \mathcal{N}(S)$ contains species $A$ and $s_i = 0$ otherwise, then the whole problem becomes that of paramagnetic spins on a network subject to an effective node-dependent magnetic field whose magnitude is $G(w)\Lambda^2(w)\Delta(w)$ and where $\Delta(w)$ denotes the euclidean area of the triangular unit at point $w$ in the complex plane. Since the effective magnetic field is non uniform and non trivial, there is no simple explicit analytical expression for the thermodynamically favoured composition morphologies. By splitting the system into 8 equivalent patches, we could however suggest, as discussed in the main text, what would happen at low enough temperatures. In particular, a first approximation scheme neglecting the effect of curvature on the area measure and Taylor expanding the curvature field about its zero point $p_i$ suggested that the curvature energy $H^{\Sigma_i}_{el}(f_A^i)$ in a patch $\Sigma_i$, $i=1..8$, with an area fraction $f_A^i$ of $A$ lipids would go as $H^{\Sigma_i}_{el}(f_A^i) \sim (f_A^i)^{\alpha}$ with $\alpha = 2$. To test numerically this proposition, we performed Monte Carlo (MC) simulations of an Ising system whereby: 
\begin{enumerate}
\item The total number of spins/cells is fixed,
\item The total number of spins of value 1 is fixed,
\item A MC move consists then in:
\begin{enumerate}
\item  picking at random a cell among those which have $s=1$, 
\item  picking at random a cell among those which have $s=0$,
\item  swap the cells spin values,
\item  accept the move with a probability satisfying the Metropolis criterion \cite{Metropolis} $p_{acc} = \min[1, e^{-\beta \Delta E}]$, where $\Delta E = E_{final}-E_{initial}$ and the energy is in general given by Eqs. (4) and (5) of the main text article.
\end{enumerate}
\end{enumerate}
If we set the Ising parameter $J$ to zero, we can then probe the low energy curvature energy as a function of domain size for a single patch.
\begin{figure}
 \vspace{2mm}
  \includegraphics[width=0.90\columnwidth]{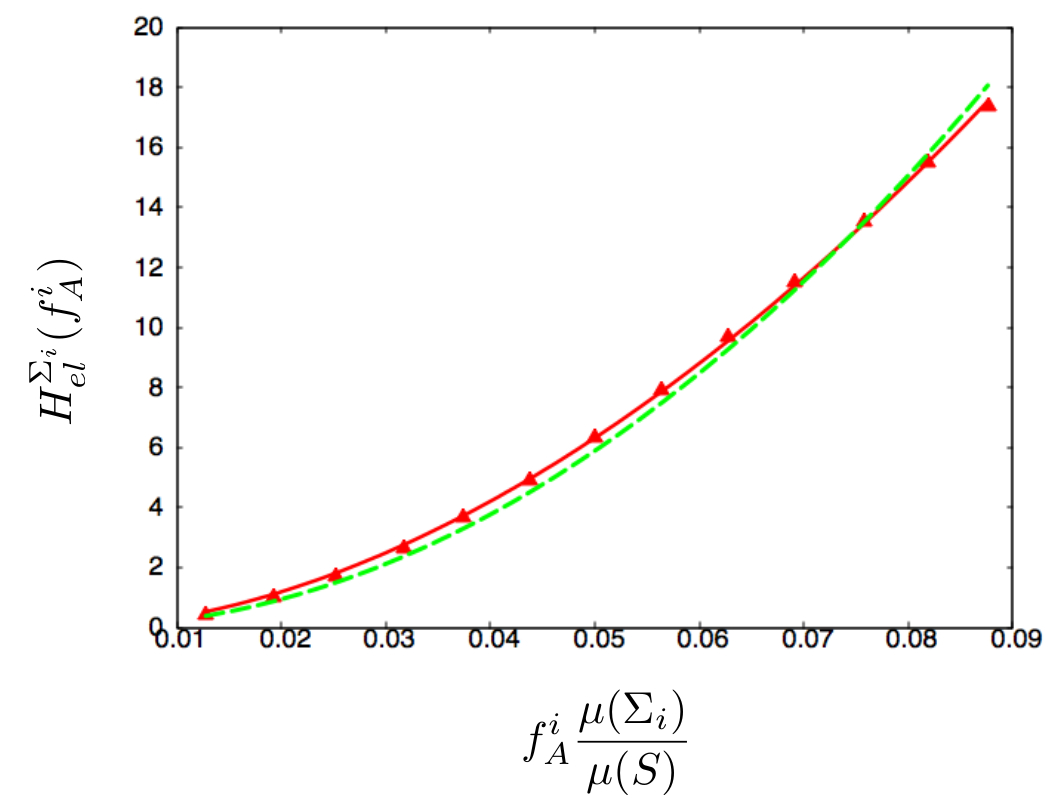} 
  
  \caption{\label{fig2} {\it Curvature energy as a function of domain size}. The red triangles are MC data points for the energy $H^{\Sigma_i}_{el}(f_A^i)$ of an $A$ domain on a patch $\Sigma_i$. The lines correspond to the best fit to these data points with a scaling law behaviour $H^{\Sigma_i}_{el}(f_A^i) \sim (f_A^i)^{\alpha}$. The solid red line corresponds to the best fit exponent value $\alpha \approx 1.83$, while the green dashed line is the best fit to the data with imposed $\alpha = 2$ which shows quite good agreement with the data.}
\end{figure}As we see in Fig. \ref{fig2}, the proposition that the curvature energy goes as a power low is in very good agreement with MC data. In addition, the approximation that $\alpha = 2$ is in remarkably quite good agreement with the data.

\section{Validity of the minimal surface assumption}
In the current study, it is assumed that the underlying surface remains minimal during segregation of the lipids/species. However, in general domain formation can induce local membrane deformation which in turn can destabilise the entire morphology of the membrane itself. In this section we will have a closer look at how phase separation induced bud formation may affect the conclusions drawn in the manuscript.

\subsection{Budding on flat multicomponent membranes}
Based on the work by Lipowsky \cite{Lipowsky1992} on flat membranes, domain formation may lead to a budding phenomenon driven by the line tension between the two coexisting demixed phases. Budding occurs when the line tension energy cost overcomes the bending energy penalty. 

Consider a membrane domain of area $A = \pi L^2 = 2\pi R^2 (1-\cos\theta)$, where $R$ is the radius of curvature of the deformed membrane domain, $\theta$ is the contact angle of the domain with respect to the horizontal plane, and
$L$ is the domain radius if there is no deformation to the membrane. 
We will now consider the competition between two energy terms: a) line tension energy that increases with the perimeter of the domain, $2\pi J R \sin \theta$; and (ii) bending energy which depends on the domain area and curvature, $2 \kappa_m A / R^2$.
Here $J$ is the line tension and $\kappa_m$ is the mean curvature bending rigidity. We have also assumed that there is no mean spontaneous curvature. 

\begin{figure}[h]
 \vspace{2mm}
  \includegraphics[width=0.99\columnwidth]{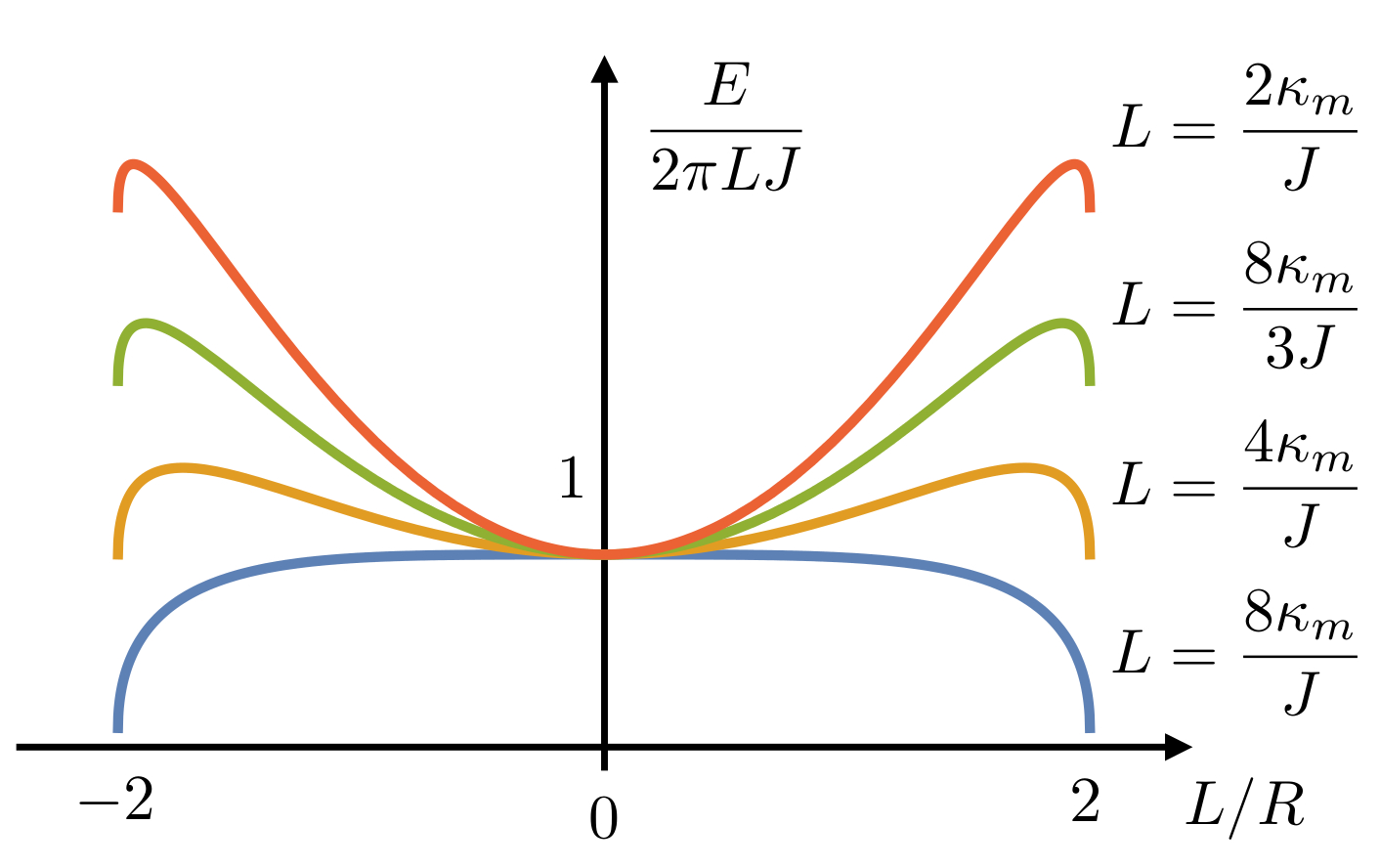} 
    \caption{\label{fig7} {\it Sum of the line tension and curvature energies of a membrane domain}. Normalised energy curves as a function of reduced curvature $L/R$ for given values of the domain size $L$. From top to bottom the domain size $L$ is increased from $L^*/2$ to $L^o$.  }
\end{figure}

Fig. \ref{fig7} shows the total energy (normalised by the line tension energy for a flat domain) as a function of the reduced membrane mean curvature $L/R $ that plays the role of an order parameter. $L/R = 0$ corresponds to a flat membrane, i.e. no deformation. $L/R = 2$ corresponds to a complete bud formation. As we see, there are three regimes depending on the membrane domain size $L$: (i) For $0 < L <  L^*= 4\kappa_m/J $, budding is unfavourable and $L/R = 0$ is the global minimum configuration; (ii) For $ L^* \le L <  L^o = 8\kappa_m/J $, budding is favourable but there is an energy barrier for its formation; (iii) Finally, only for $ L \ge L^o$ that the energy barrier for bud formation disappears. Here the flat membrane geometry is completely unstable.

\vspace{2mm}
%

\subsection{Bud formation on a P-surface}
Since the P-surface is a minimal surface with zero mean curvature everywhere, we will continue to assume that the spontaneous mean curvature is zero, as in the previous paragraph.
The presence of non-uniform gaussian curvature on the P-surface may alter the numerical prefactors in $L^*$ and $L^o$. 
However, since the domain formation occurs in the neighbourhood of very specific points on the P-surface {\it i.e.} 8 zero-curvature points each at the centre of a patch in the cubic cell, to first approximation, it is reasonable to assume that the above results from \cite{Lipowsky1992} to hold. As a result, we can approximate the two critical domain sizes as $L^* = 4\kappa_m/J$ and $L^o = 8 \kappa_m/J$.
Correspondingly, when $L < L^*$, we expect our assumption that the underlying surface remains minimal during phase separation to hold. Strong deviations to the results presented in the main text are only expected when $L \ge L^*$.

\begin{figure}[h]
 \vspace{2mm}
  \includegraphics[width=0.99\columnwidth]{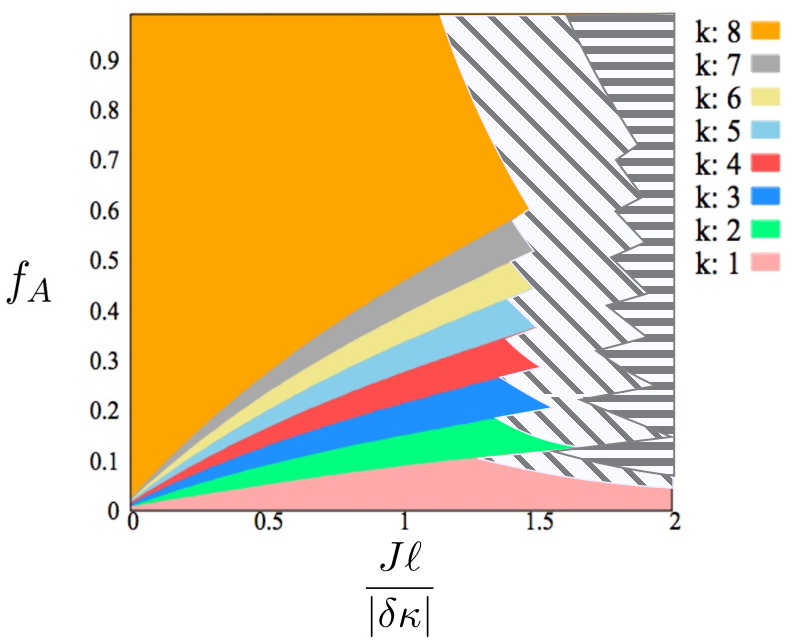} 
  
  \caption{\label{fig8} Modification of the phase diagram of Fig. 2(a) in the main manuscript resulting from budding instability for $\kappa_m/|\delta \kappa| = 1/4$. The diagonally hashed region corresponds to a region where budding is preferable but there is an energy barrier for its formation, while the horizontally hashed region corresponds to a fully unstable bud formation. }
\end{figure}

To see which part of the phase diagram in Fig. 2 of the main manuscript is affected by the budding instability, we relate the area fraction $f_A$ occupied by the A-lipids to the size $L$ of the domains depending on which configuration $\binom 8 k $ they are in. For the purpose of this analysis, we will focus on cases where there are no bridge formations between lipid $A$ domains. The total area occupied by the A-lipids on a cubic cell of the P-surface is $f_A \mu_S(S)$. If the A-lipids are partitioned in $k$ patches among the 8 available, then the typical area per domain is $f_A \mu_S(S)/k \approx \pi L^2$. It follows that requiring full stability against budding ($L < L^*$) is equivalent to requiring $f_A \mu_S(S)/k < \pi 16 (\kappa_m/J)^2$. If the cubic cell bounding the P-surface has sides of length $\ell$, then $\mu_S(S) = 24 \ell^2 K(1/4)/K(3/4)$, where $K(x)$ is the complete elliptic integral of the first kind \cite{Gandy-P}. Thus, the stability criterion becomes
\begin{equation}
f_A^*\left(\frac{J \ell}{|\delta \kappa|}; k\right) = \left( \frac{\kappa_m}{|\delta \kappa|} \right)^2 \frac{2 k \pi K(3/4)}{3K(1/4)} \left(\frac{|\delta \kappa|}{J \ell}\right)^2  \tag{SM21} \label{criticallines1}
\end{equation}
for configuration $\binom 8 k $.
As we see in Eq. \eqref{criticallines1}, the set of stability lines depends on the ratio $\kappa_m /|\delta \kappa|$ which constitutes an additional parameter in our modelling. We have observed that for values $\kappa_m/|\delta \kappa| > 0.4$, there is no intersection between any of the stability lines and the lipid repartition phases they correspond to within the parameter ranges of the present study, $0 \le J \ell / |\delta \kappa| \le 2$. 

It is worth emphasizing that experimental and numerical estimates of the ratio $\kappa_m/|\delta \kappa|$ for various lipid systems \cite{Hu12} show that it is rarely below 1. As an example, consider the mixture of DOPC, sphingomyelin, and cholesterol forming coexisting $L_o$ and $L_d$ domains reported in references \cite{Semrau,Baumgart}. Here the difference in Gaussian bending moduli $|\delta \kappa| \simeq 3 \times 10^{-19}$ J and the mean curvature bending modulus for the $L_o$ phase $\kappa_m \simeq 8 \times 10^{-19}$ J, which gives us $\kappa_m/|\delta \kappa| \simeq 8/3$. Even if we use the mean curvature bending modulus for the $L_d$ phase $\kappa_m \simeq 2 \times 10^{-19}$ J, which may be more appropriate when bridges are formed such that the lipid B domains are now surrounded by A (see Fig. 2(b) and (c) in the main text), we still have $\kappa_m/|\delta \kappa| \simeq 2/3 > 0.4$. Thus, this strongly suggests that our conclusions in the main text will hold even when taking into account segregation induced membrane deformation.

We can redo a similar calculation to that leading to Eq. \eqref{criticallines1} to determine the phase boundaries beyond which the membrane is completely unstable against bud formation. Not surprisingly we find that they satisfy a very similar equation:
\begin{equation}
f_A^o\left(\frac{J \ell}{|\delta \kappa|}; k\right) = \left( \frac{\kappa_m}{|\delta \kappa|} \right)^2 \frac{4 k \pi K(3/4)}{3K(1/4)} \left(\frac{|\delta \kappa|}{J \ell}\right)^2. \tag{SM22} \label{criticallines2}
\end{equation}

To illustrate how the phase diagram is modified when budding instability is taken into account, we show the results for $\kappa_m /| \delta \kappa| = 1/4$ in Fig. \ref{fig8}. The parameter regime susceptible to budding is for large area fraction of $A$ lipids, $f_A$, and large line tension, $J \ell/|\delta \kappa|$. The hashed regions correspond  to parameter regimes where budding is preferable. Energy barriers are present for budding to occur in the diagonally hashed region, while for the horizontally hashed region the P-surface is completely unstable against budding.

\bibliographystyle{apsrev4-1}
\bibliography{biblio}

\end{document}